\documentclass[prd,showpacs,amsmath,amssymb,superscriptaddress,preprintnumbers,nofootinbib,showkeys]{revtex4}

\begin{document}

\newcommand{\nablab}{{\mathop {\rule{0pt}{0pt}{\nabla}}\limits^{\bot}}\rule{0pt}{0pt}}

\title{Extended relativistic non-equilibrium thermostatics \\ of stellar structures with radiation pressure}

\author{Alexander B. Balakin}
\email{Alexander.Balakin@kpfu.ru} \affiliation{Department of
General Relativity and Gravitation, Institute of Physics, Kazan
Federal University, Kremlevskaya str. 16a, Kazan 420008, Russia}

\author{Zagir Z. Tukbaev}
\email{zagir2324@gmail.com} \affiliation{Department of
General Relativity and Gravitation, Institute of Physics, Kazan
Federal University, Kremlevskaya str. 16a, Kazan 420008, Russia}

\date{\today}

\begin{abstract}
We establish the extended formalism for description of the static spherically symmetric relativistic non-equilibrium stellar systems in the formation of which the radiation pressure plays the key role. The main concept of this extended formalism inherits the ideas, on which the Israel-Stewart causal thermodynamics is based, but now the unit spacelike four-vector, indicated by the term {\it director}, is exploited in addition to the unit timelike medium velocity four-vector. An application of the extended formalism is considered; we analyze the profiles of the non-equilibrium pressure and temperature as the functions of guiding parameters introduced phenomenologically.
\end{abstract}
\maketitle

\section{Introduction}

In the heart of the Milky Way the supermassive black hole was detected \cite{1BH}. The publication of the Science Release ESO1825 concerning this event has become the culmination of long-term observations, and has opened a new page in the upgrade of the theory of evolution of stellar structures (see, e.g.,\cite{2BH,3BH}). One of the zones of the  Hertzsprung-Russell diagram, which attracts the interest in this context, is the area of stellar structures, in the evolution of which the radiation pressure plays an essential role. What contribution to the development of the star evolution science the theoreticians could make? We assume that one of the most interesting trends in this direction is the non-local rheological-type extension of the relativistic non-equilibrium irreversible thermodynamics and thermostatics. In the work \cite{BalNS} we have made the step towards the development of the theory based on rheological-type extension of the equation of state of the neutron stars at zero temperature. In this work we consider the stellar objects with high temperature and assume that the radiation pressure is the key player in the corresponding equation of state. We establish our extended model based on the ideas of causal irreversible thermodynamics elaborated by Israel and Stewart \cite{IsraelStewart}; the short prologue about the mathematical formalism of this theory is presented in Section II.

We would like to emphasize one detail in the difference between the extended thermodynamics
\cite{JCL} and extended relativistic thermostatics. The
extended thermodynamics
 deals with the heat propagation. The corresponding equation for the temperature evolution is hyperbolic (of the second order in derivative with respect to time) in both versions: proposed by Cattaneo \cite{Cattaneo} and resulting from the Israel-Stewart theory \cite{IsraelStewart}. This result is due to accounting for the retardation of the response, the simplest manifestation of the non-locality in time. The extended relativistic thermostatics does not operate with time derivatives and thus has to exploit the idea of spatial non-locality. Using this theory one analyzes the static temperature distribution instead of temperature evolution, and the mathematical formalism has to be correspondingly extended. We describe these modifications of the formalism in Section III.

The last remark is the following. The causal thermodynamics was used in many works for the analysis of the rate of cosmological expansion, of the dynamics of perturbations, etc. (see, e.g., \cite{CT1,CT2,CT3,CT4,CT5,CT6}). When the system is static, one deals most often with the relativistic thermostatics of spherically symmetric bodies, and the equation of hydrostatic equilibrium becomes the central element of the analysis. We also considered the static system with the spherical symmetry, and arranged the results of analysis in Section IV. Section V contains conclusions.

\section{Prologue: The canonic causal relativistic non-equilibrium thermodynamics}

The story of irreversible relativistic thermodynamics  has a remarkable page associated with the so-called causal thermodynamics elaborated by Israel and Stewart \cite{IsraelStewart}. This theory is based on the second law of the phenomenological thermodynamics, which states that the entropy production $\sigma$ of a closed physical system should be non-negative $\sigma \geq 0$. Entropy production scalar is introduced as the covariant divergence of the entropy flux four-vector $S^k$, i.e., $\sigma=\nabla_kS^k$ ($\nabla_k$ is the covariant derivative). Modeling of the vector $S^k$ is the crucial point of the corresponding theory. In order to motivate the proposed extension of causal thermodynamics we would like to recall shortly the main details of this theory.

\subsection{Eckart's approach}

According to the Eckart version of linear thermodynamics \cite{Eckart} the entropy flux four-vector has the form
\begin{equation}
S^k_{(\rm Eckart)} = s_0 n U^k + \frac{1}{T} q^k \,,
\label{1}
\end{equation}
where $n$ is the scalar of particle number density, $T$ is the temperature, $s_0$ is the scalar of entropy per one particle, $U^k$ is the unit timelike medium velocity four-vector, and $q^k$ is the spacelike heat-flux four-vector.
The scalar $s_0$ enters the Gibbs equation (the first law of thermodynamics)
\begin{equation}
\delta e + P \delta \left(\frac{1}{n} \right) = T \delta s_0 \,,
\label{02}
\end{equation}
where $e$ describes the energy density per one particle, and $W=en$ is the energy density; $P$ is the isotropic equilibrium Pascal pressure, and the
operator $\delta$ is connected with the variation of the corresponding thermodynamic quantity. In the Eckart approach the symbol $\delta$ is replaced with the
differential operator $D$, the convective derivative defined as $D=U^k \nabla_k$ . The stress-energy tensor of the medium can be decomposed standardly as follows:
\begin{equation}
T^{ik} = en U^i U^k + U^i q^k + U^k q^i - \Delta^{ik} P + \Pi^{ik} \,.
\label{03}
\end{equation}
Here $\Delta^{ik} = g^{ik}{-}U^iU^k$ is the projector, and the tensor
\begin{equation}
\Pi^{ik} = \Pi^{ik}_{(0)} + \frac13 \Delta^{ik} \Pi \,, \quad \Pi \equiv g_{ik} \Pi^{ik} \,,
\label{04}
\end{equation}
describes the non-equilibrium pressure of the medium. The stress-energy tensor is considered to be divergence - free, $\nabla_k T^{ik}=0$, as usual, we split these four equations into the scalar and vector subsets, respectively:
\begin{equation}
D W + (W+P)\Theta - q^k DU_k + \nabla_k q^k -\Pi^{ik} \nabla_kU_i= 0 \,,
\label{05}
\end{equation}
\begin{equation}
(W+P) DU^j - \nablab^j P =
- q^j \Theta - q^k \nablab_k U^j - \Delta^j_k Dq^k + \Pi^{jk} DU_k -\Delta^j_m \nablab_k \Pi^{mk} \,.
\label{06}
\end{equation}
The quantity $\Theta = \nabla_kU^k$ is the scalar of expansion of the medium flow; the operator $\nablab_k = \Delta_k^l \nabla_l$ plays the role of spatial part of the gradient. It is well known that, when one uses the Gibbs equation (\ref{02}) plus the energy conservation law (\ref{05}), the entropy production scalar is calculated to have the form
\begin{equation}
\sigma= \frac{1}{T} \left[\Pi^{ik} \nablab_k U_i + q^k \left(DU_k - \frac{1}{T} \nablab_k T\right) \right] \,,
\label{7}
\end{equation}
where $DU_k$ is the acceleration four-vector. According to the Eckart results, the entropy production scalar $\sigma$ is non-negative, when
\begin{equation}
q^i = \lambda \left[\nablab^i T {-} T DU^i \right], \quad \Pi = 3 \zeta \Theta  \,, \quad
\Pi_{ik(0)} {=} \eta \sigma_{ik} \,,
\label{8}
\end{equation}
since such phenomenological ansatz guarantees  that
\begin{equation}
T \sigma = - \frac{1}{\lambda T} q^k q_k + \frac{1}{\eta} \Pi^{ik}_{(0)} \Pi_{ik (0)} + \frac{1}{9\zeta} \Pi^2 \,,
\label{9}
\end{equation}
and the the entropy scalar to be non-negative $\sigma \geq 0$. The phenomenological constants: $\lambda$ (the thermal conductivity), $\eta$ (the shear viscosity) and $\zeta$ (the bulk viscosity), are assumed to be functions of the temperature $T$. In (\ref{8}) the standard elements of the decomposition of the velocity covariant derivative are used:
\begin{equation}
\nabla_m U_n = U_m D U_n + \sigma_{mn} + \omega_{mn} + \frac13 \Delta_{mn} \Theta \,,
\label{10}
\end{equation}
where the symmetric traceless shear tensor $\sigma_{mn}$  and the skew-symmetric vorticity tensor $\omega_{mn}$ are given, respectively, as follows:
\begin{equation}
\sigma_{ik} \equiv \left[ \frac12 \left(\nablab_i U_k + \nablab_k U_i \right) - \frac13 \Theta \Delta_{ik} \right] \,, \quad
\omega_{mn} \equiv \frac12 \left(\nablab_i U_k - \nablab_k U_i \right) \,.
\label{1277}
\end{equation}
These tensors are orthogonal to the velocity four-vector $U^k$, i.e., $\sigma_{mn} U^m =0=\sigma_{mn} U^n$ and $\omega_{mn} U^m =0=\omega_{mn}U^n$.

\subsection{Approach of Israel and Stewart}

In the framework of causal thermodynamics Israel and Stewart have used the following ansatz for the entropy flux four-vector structure:
\begin{equation}
S^k_{(\rm IS)} = S^k_{(\rm Eckart)} + \frac{1}{T} q^l \left[\delta^k_l \alpha_0 \Pi  + \alpha_1 \Pi^k_{\ l(0)} \right] -
\frac{1}{2T} U^k \left[\beta_0 \Pi^2 - \beta_1 q^m q_m + \beta_2 \Pi^{mn}_{(0)} \Pi_{mn(0)}\right] \,.
\label{12}
\end{equation}
In other words, the authors of this version of the causal thermodynamics have added all the possible terms of the second order with respect to the non-equilibrium quantities $\Pi$, $q^k$ and
$\Pi_{mn(0)}$ with new phenomenological parameters $\alpha_0$, $\alpha_1$, $\beta_0$, $\beta_1$ and $\beta_2$. Using the same scheme of calculation of the entropy production scalar, as in the Eckart version, one can obtain the formula
$$
T \sigma = \Pi \left[\frac13 \Theta - \beta_0 D \Pi - \frac12 T \Pi \nabla_l \left(\frac{\beta_0 U^l}{T} \right) + \alpha_0 \nabla_k q^k \right] +
$$
$$
+ q^k \left[DU_k - \frac{1}{T} \nablab_k T + \beta_1 D q_k  + \frac{\beta_1}{2}  q_k \Theta
+ \frac12 T q_k D \left(\frac{\beta_1}{T} \right) + T \nabla_k \left(\frac{\alpha_0 \Pi}{T} \right) + T \nabla_l \left(\frac{\alpha_1}{T} \Pi^l_{\ k (0)}\right)\right] +
$$
\begin{equation}
+ \Pi^{ik}_{(0))} \left[\sigma_{ik} {-} \beta_2 D \Pi_{ik (0)} {-} \frac{\beta_2}{2} \Pi_{ik (0)} \Theta  -
\frac12 T \Pi_{ik (0)} D \left(\frac{\beta_2}{T} \right) + \alpha_1 \nabla_{(i} q_{k)} \right] \,.
\label{13}
\end{equation}
The entropy production scalar $\sigma$ can be presented again as a non-negative quantity (\ref{9})
if one uses the following definitions for $\Pi$, $q^i$ and $\Pi_{ik (0)}$:
\begin{equation}
\beta_0 D \Pi {+} \Pi \left[\frac{1}{9\zeta} {+} \frac{T}{2} (\Theta {+}D) \left(\frac{\beta_0}{T} \right)\right] = \frac13 \Theta {+} \alpha_0 \nabla_k q^k   \,,
\label{14}
\end{equation}
\begin{equation}
\beta_1 \Delta^k_l D q^l + q^k \left[\frac{1}{\lambda T} +  \frac{T}{2}\left(\Theta+D \right)\left(\frac{\beta_1}{T}\right) \right] =
\frac{1}{T} \nablab^k T {-} DU^k  {-} T \nablab^k \left(\frac{\alpha_0 \Pi}{T} \right) {-} T \Delta^{k}_{s} \nabla_l \left(\frac{\alpha_1}{T} \Pi^{ls}_{(0)}\right) \,,
\label{15}
\end{equation}
\begin{equation}
\beta_2 \Delta^m_i \Delta^n_k D \Pi_{mn (0)} {+} \Pi_{ik (0)} \left[\frac{1}{\eta} {+}  \frac{T}{2}\left(\Theta {+} D \right) \left(\frac{\beta_2}{T}\right)\right]
= \sigma_{ik}  + \alpha_1 \Delta^m_i \Delta^n_k \nabla_{(m} q_{n)} \,.
\label{16}
\end{equation}
This canonic result shows that the expansion scalar $\Theta$ is the source of the non-equilibrium pressure scalar $\Pi$; the shear tensor $\sigma_{ik}$ is the source of the quantity $\Pi^{ik}_{(0)}$; the difference $\frac{1}{T} \nablab^k T {-} DU^k$ is the source of the heat flux. If these sources vanish, there exists the trivial solutions for the mentioned non-equilibrium fluxes.
The phenomenological parameters $\beta_0$, $\beta_1$, $\beta_2$ predetermine the rates of evolution of the corresponding  non-equilibrium fluxes.

\section{Extension of the relativistic non-equilibrium thermostatics}

\subsection{Three remarks concerning the thermostatics of the objects with spherical symmetry}

\subsubsection{The structure of covariant derivative of the velocity four-vector}

The canonic theory of static spherically symmetric stellar structures is presented in the book \cite{Weinberg}.
Following this work in the whole, we nevertheless, change the signature of the metric and use the interval
\begin{equation}
ds^2 = B(r)dt^2 - A(r)dr^2 - r^2\left({d\theta}^2 + \sin^2{\theta} {d\varphi}^2 \right) \,.
\label{metric}
\end{equation}
This spacetime admits the existence of the following Killing vectors:
$$
 \xi^j_{(0)} = \delta_0^j   \,, \quad \xi^j_{(1)} = \sin{\varphi}\delta_{\theta}^j + \cot{\theta} \cos{\varphi} \delta^j_{\varphi}\,,
$$
\begin{equation}
\xi^j_{(2)} = \cos{\varphi}\delta_{\theta}^j - \cot{\theta} \sin{\varphi} \delta^j_{\varphi}\,, \quad  \xi^j_{(\varphi)} = \delta_{\varphi}^j \,.
\label{r30}
\end{equation}
We indicate the state of the physical system as inheriting the symmetry of the spacetime, when the Lie derivatives of all state functions along all the Killing vectors vanish.
The Lie derivatives of the scalars $W$, $P$, $n$, $\Pi$ vanish, i.e.,
\begin{equation}
{\cal L}_{\xi_{(a)}} W = \xi^k_{(a)}  \partial_k W  =0 \,, \quad {\cal L}_{\xi_{(a)}} P = \xi^k_{(a)}  \partial_k P =0 \,, ...
\label{r37}
\end{equation}
when these scalars depend on the radial variable $r$ only.
The velocity four-vector has to be chosen now as
\begin{equation}
U^i = \delta^i_0 \frac{1}{\sqrt{B}} \,, \quad U_i = \delta_i^0 \sqrt{B} \,.
\label{18}
\end{equation}
The covariant derivative of the velocity four-vector is now of the form
\begin{equation}
\nabla_k U_i = - \delta^0_k \delta^r_i \frac{B^{\prime}}{2\sqrt{B}}  \,.
\label{19}
\end{equation}
This means that
\begin{equation}
\Theta=0 \,, \quad \sigma_{ik}=0 \,, \quad \omega_{ik}=0 \,, \quad DU_i = -  \delta^r_i \frac{B^{\prime}}{2B} \,.
\label{20}
\end{equation}
In other words, the acceleration four-vector is the only non-vanishing object, and it has the only radial component.

\subsubsection{Is the heat-flux four-vector vanishing?}

The heat-flux four-vector $q^k$ is orthogonal to the velocity four-vector, $q^k U_k=0$, and for the ansatz (\ref{18}) we obtain that only $q^r$, $q^{\theta}$ and $q^{\varphi}$ could be nonzero. If we apply the requirement of the symmetry inheritance to the heat-flux four-vector:
\begin{equation}
{\cal L}_{\xi_{(a)}}q^i = \xi^k_{(a)} \partial_k q^i - q^k \partial_k \xi^i_{(a)} = 0 \,,
\label{r27}
\end{equation}
for all Killing vectors (\ref{r30}) we conclude, that $q^{\theta}=0$, $q^{\varphi}=0$ and $q^r$ has to depend on the radial variable only. However, in the static model the energy density balance equation (\ref{05}) converts into
\begin{equation}
0= q^k DU_k - \nabla_k q^k  \ \Rightarrow \  - q^r \frac{B^{\prime}}{2B} = \frac{1}{r^2\sqrt{AB}}\frac{d}{dr}\left(r^2\sqrt{AB} q^r \right) \,,
\label{057}
\end{equation}
thus, the solution to (\ref{057}) is
\begin{equation}
q^r = \frac{K}{r^2 B\sqrt{A}} \,.
\label{058}
\end{equation}
Physical motives hint us that we have to put the integration constant $K$ to zero, and thus, we have to use the ansatz $q^i=0$. Then
the equation for $\Pi$ (\ref{14}) becomes sourceless and thus (\ref{14}) prescribes the scalar non-equilibrium pressure to vanish, $\Pi{=}0$. Similarly, since the shear tensor vanishes, $\sigma_{ik}=0$,  the tensor $\Pi_{mn (0)}$ turns into zero. Finally, we obtain from (\ref{15}), that
\begin{equation}
q^i=0 \ \Rightarrow  \  \frac{1}{T} \nablab_k T = DU_k \,,
\label{21}
\end{equation}
and taking into account (\ref{20}) we obtain immediately, that
\begin{equation}
T(r)= \frac{T_0}{\sqrt{B(r)}} \,.
\label{22}
\end{equation}
In other words, for the static spherically symmetric models all the non-equilibrium  fluxes vanish
\begin{equation}
\Pi=0 \,, \quad  q^i =0 \,, \quad \Pi_{mn (0)} =0 \,,
\label{23}
\end{equation}
and the distribution of the temperature is described by the equilibrium law (\ref{22}). This means, in fact, that the canonic Israel-Stewart theory is not effective in the case, when we study the static spherically symmetric stellar configurations, and we have to think about the extension of the causal thermostatics.

\subsubsection{Special solution for the heat-flux four-vector}

Formally speaking, we can consider the case $K \neq 0$ also. Let us assume that $\alpha_1=0$, thus $\Pi^{ik}_{(0)}$ has no physical source and we put $\Pi^{ik}_{(0)}=0$.
Then, the equations (\ref{05}), (\ref{14}) and (\ref{15}) yield, respectively,
\begin{equation}
q^k DU_k = \nabla_k q^k  \ \Rightarrow q^k = \delta^k_r \frac{K}{r^2B\sqrt{A}}\,,
\label{143}
\end{equation}
\begin{equation}
\frac{\Pi}{9\zeta} = \alpha_0 \nabla_k q^k  \ \Rightarrow \ \Pi = - \frac{9 \zeta \alpha_0 K B^{\prime}}{2r^2B^2\sqrt{A}} \,,
\label{144}
\end{equation}
\begin{equation}
\frac{q_k }{\lambda T} =
\frac{1}{T} \nablab_k T {-} DU_k  {-} T \nablab_k \left(\frac{\alpha_0 \Pi}{T} \right) \Rightarrow
T^{\prime}\left(1+\alpha_0 \Pi \right) + T \left[\frac{B^{\prime}}{2B} - \left(\alpha_0 \Pi \right)^{\prime} \right] = - \frac{K\sqrt{A}}{\lambda r^2 B}
 \,.
\label{151}
\end{equation}
In other words, when $K \neq 0$, the equations (\ref{143})-(\ref{151}) describe special branch of solutions for the thermostatic model. In particular, if we choose $\alpha_0=0$ and $\lambda(r) = \frac{\lambda_0 \sqrt{A}}{B}$ with the constant $\lambda_0$, the solution to the equation (\ref{151}) is
\begin{equation}
T(r) = \frac{T_0}{\sqrt{B}}\left(1+ \frac{K}{\lambda_0 T_0 r} \right) \,.
\label{c1}
\end{equation}
There is one interesting case, when $K$ is negative: one can denote $r_{\rm C}= \frac{|K|}{\lambda_0 T_0}$ and obtain $T(r) = \frac{T_0}{\sqrt{B}}\left(1-\frac{r_{\rm C}}{r} \right)$.
Thus the constant of integration $K$ can be linked with the critical value of the radial variable $r_{\rm C}$: when $r<r_{\rm C}$, the temperature becomes negative, and the thermostatic description becomes inappropriate.

\subsection{The director and its properties}

If we follow the idea that in the framework of thermostatics of the spherically symmetric bodies the heat-flux four-vector $q^k$ disappears (i.e., $K=0$), we need of new four-vector orthogonal to the velocity four-vector $U^j$.
We suggest to use the spacelike unit four-vector ${\cal R}^i$, which inherits the symmetry of the spacetime. This means that
\begin{equation}
{\cal R}^i U_i =0 \,, \quad {\cal R}_i {\cal R}^i = -1 \,,
\label{r1}
\end{equation}
and the Lie derivatives along all the Killing vectors (\ref{r30}), admitted by the spacetime symmetry, vanish:
\begin{equation}
{\cal L}_{\xi_{(a)}}{\cal R}^i = \xi^k_{(a)} \partial_k {\cal R}^i - {\cal R}^k \partial_k \xi^i_{(a)} = 0 \,.
\label{r2}
\end{equation}
All the requirements (\ref{r1}), (\ref{r2}) are satisfied for the four-vector
\begin{equation}
{\cal R}^i = \delta^i_r \frac{1}{\sqrt{A}} \,, \quad {\cal R}_i = - \delta_i^r {\sqrt{A}} \,.
\label{r5}
\end{equation}
The covariant derivative of this vector
\begin{equation}
\nabla_k {\cal R}^i = \frac{1}{2\sqrt{A}}\left[\frac{B^{\prime}}{B} \delta^i_0 \delta_k^0 + \frac{2}{r} \left( \delta^i_{\theta} \delta_k^{\theta} +\delta^i_{\varphi} \delta_k^{\varphi} \right) \right]  \,,
\label{r6}
\end{equation}
can be rewritten as
\begin{equation}
\nabla_k {\cal R}^i = \frac{B^{\prime}}{2B \sqrt{A}} U^i U_k  + \frac{1}{r\sqrt{A}} \left( \delta^i_{\theta} \delta_k^{\theta} +\delta^i_{\varphi} \delta_k^{\varphi} \right)   \,.
\label{r63}
\end{equation}
Clearly, the tensor $\nabla_k {\cal R}_j$ is symmetric, and its trace is equal to
\begin{equation}
\nabla_k {\cal R}^k = \frac{1}{2\sqrt{A}}\left[\frac{B^{\prime}}{B} + \frac{4}{r} \right]  \,.
\label{r7}
\end{equation}
Using the director ${\cal R}^k$, one can rewrite the relationship (\ref{19}) in more convenient form
\begin{equation}
\nabla_k U^i =  \frac{B^{\prime}}{2B\sqrt{A}} U_k {\cal R}^i   \,.
\label{191}
\end{equation}
Since $U^k$ and ${\cal R}^k$ are orthogonal, we see explicitly that $\nabla_k U^k =0$.
Similarly, taking into account (\ref{r6}) and (\ref{191}), as well as the normalization conditions for the four-vectors $U^k$ and ${\cal R}^j$, we can write the system of useful relationships
\begin{equation}
{\cal R}^k \nabla_k {\cal R}^i = 0 \,,  \quad {\cal R}^k \nabla_k U^i =0 \,,
\quad U_i \nabla_k U^i = 0 \,,  \quad {\cal R}_i \nabla_k {\cal R}^i = 0 \,,
\label{r94}
\end{equation}
\begin{equation}
U^k \nabla_k {\cal R}^i = \Gamma U^i \,, \quad U_i \nabla_k {\cal R}^i = \Gamma U_k \,,
\quad U^k \nabla_k U^i = \Gamma {\cal R}^i \,,  \quad {\cal R}_i \nabla_k U^i = - \Gamma U_k   \,,
\label{r92}
\end{equation}
where we introduced the following auxiliary function
\begin{equation}
 \Gamma(r) \equiv \frac{B^{\prime}}{2B\sqrt{A}} \,.
\label{r93}
\end{equation}
Clearly, there exists some symmetry in these relationships between the medium velocity $U^i$ and the director ${\cal R}^j$. It is interesting to mention that in all formulas (\ref{r92}) the common  multiplier $\Gamma$ (\ref{r93}) appeared.

\subsection{Extension of the entropy flux four-vector and modified equations for the  pressure $\Pi$ and temperature}

We assume now that $q^i$ and $\Pi_{(0)}^{ik}$ do not participate in the extension  procedure; only $\Pi$ appears in the new terms of the decomposition of the entropy flux four-vector $S^i$.
We add to the decomposition (\ref{12}) the following new terms:
\begin{equation}
S^k = S^k_{(\rm IS)} + {\cal R}^k \left[ \frac12  \tau \Pi^2 + \frac13 \gamma \Pi^3 + ... \right] \,,
\label{26}
\end{equation}
where the multipliers $\tau(T,n)$ and $\gamma(T,n)$ are some functions of the temperature and particle number density, introduced phenomenologically.
The constitutive equation for the scalar $\Pi$ takes now the form
\begin{equation}
\beta_0 D \Pi {+} \Pi \left[\frac{1}{9\zeta} {+} \frac{T}{2} (\Theta {+}D) \left(\frac{\beta_0}{T} \right)\right] {-} \frac13 \Theta {-} \alpha_0 \nabla_k q^k =
T\left[\left(\tau {+} \gamma \Pi \right){\cal R}^i \nabla_i \Pi {+} \frac12 \Pi \nabla_i \left(\tau {\cal R}^i \right) {+}   \frac13 \Pi^2 \nabla_k \left(\gamma {\cal R}^k \right) \right]\,.
\label{27}
\end{equation}
Keeping in mind that the model under consideration is static and spherically symmetric, we can reduce the constitutive equation (\ref{27}) into
\begin{equation}
\left(\tau + \gamma \Pi \right){\cal R}^i \nabla_i \Pi + \frac12 \Pi \nabla_i \left(\tau {\cal R}^i \right) +  \frac13 \Pi^2 \nabla_k \left(\gamma {\cal R}^k \right) = \frac{\Pi}{9\zeta T}\,.
\label{271}
\end{equation}
In (\ref{271}) the differential operator ${\cal D} {=} {\cal R}^k \nabla_k$ appears, which plays the similar role as the operator $D {=} U^k \nabla_k$ in the equation (\ref{14}).
When the non-equilibrium pressure $\Pi$ is non-vanishing, the law of the temperature distribution (see (\ref{15})) transforms into
\begin{equation}
\frac{1}{T} \nablab_k T = DU_k  + T\nablab_k \left(\frac{\alpha_0 \Pi}{T} \right)  \,.
\label{30}
\end{equation}
In this set only one equation is nontrivial. Convolution of (\ref{30}) with ${\cal R}^k$ gives the equation
\begin{equation}
\frac{d}{dr}\left[\log{\left(T \sqrt{B}\right)} \right] = T \frac{d}{dr}\left(\frac{\alpha_0 \Pi}{T} \right) \,.
\label{31}
\end{equation}
When $\alpha_0=0$, or $\frac{\alpha_0 \Pi}{T} = const$, the solution to (\ref{31}) again is $T(r)=\frac{T_0}{\sqrt{B}}$,
where $T_0$ is a constant of integration.

\subsection{Two particular models}

\subsubsection{The model with $\gamma=0$}

When the decomposition of the entropy flux (\ref{26}) contains the quadratic terms only, i.e., when $\gamma=0$, the equation (\ref{271}) takes the form
\begin{equation}
\tau {\cal R}^i \nabla_i \Pi + \frac12 \Pi \nabla_i \left(\tau {\cal R}^i \right)  = \frac{\Pi}{9\zeta T}\,,
\label{272}
\end{equation}
or in more details (again the prime denotes the derivative with respect to radial variable)
\begin{equation}
\tau \Pi^{\prime} + \frac12 \Pi  \left[\tau^{\prime} + \tau \left(\frac{B^{\prime}}{2B} + \frac{2}{r}\right) \right]  = \frac{\Pi \sqrt{A}}{9\zeta T}\,.
\label{273}
\end{equation}
Clearly, the trivial solution $\Pi=0$ satisfies this equation. We could consider nontrivial solutions, when, e.g., $\Pi(r_*)\neq 0$ on some sphere $r=r_*$. We can rewrite the equation (\ref{273}) in the form
\begin{equation}
\frac{H^{\prime}}{H} =  \frac{\sqrt{A}}{9 \zeta T \tau} \,, \quad  H = \Pi \sqrt{\tau r^2 \sqrt{B}}\,.
\label{274}
\end{equation}
In the particular case, when $\zeta(T){=} \frac{\zeta_0}{T}$, and $\tau(r) {=} \tau_0\sqrt{A}$ with constants $\zeta_0$ and $\tau_0$, we obtain the analytic solution
\begin{equation}
\Pi(r) = \Pi(r_*) \left(\frac{r_*}{r}\right)  \left(\frac{A(r_*)B(r_*)}{A(r)B(r)}\right)^{\frac14}  \ \exp{\left(\frac{r-r_*}{9\zeta_0 \tau_0}\right)}\,.
\label{275}
\end{equation}
Mention should be made that the sign of the phenomenological parameter $\tau_0$ is not yet fixed.

\subsubsection{The model with $\tau=0$, $\gamma \neq 0$}

This case is interesting since the constitutive equation for the non-equilibrium pressure $\Pi$
\begin{equation}
\Pi \left[\gamma  {\cal R}^k \nabla_k \Pi + \frac13 \Pi \nabla_k \left(\gamma {\cal R}^k \right) - \frac{1}{9\zeta T} \right] =0
\label{28}
\end{equation}
splits into the pair of equations: one of them gives the trivial solution, and the second is the equation with the source provided by the bulk viscosity
\begin{equation}
\gamma  {\cal R}^k \nabla_k \Pi + \frac13 \Pi \nabla_k \left(\gamma {\cal R}^k \right) = \frac{1}{9\zeta T} \,.
\label{29}
\end{equation}
The equation (\ref{29}) for $\Pi$ can be written as follows:
\begin{equation}
\frac{1}{r^2} \frac{d}{dr}\left[\Pi \left(r^2 \gamma \sqrt{B}\right)^{\frac13} \right] = \frac{\sqrt{AB}}{9\zeta T} \left(r^2 \gamma \sqrt{B}\right)^{-\frac23} \,.
\label{33}
\end{equation}
In particular, if $\gamma(r)=\frac{\gamma_0}{\sqrt{B}}$ and $\zeta(r)= \zeta_0 \frac{\sqrt{AB}}{T}$ with the constants $\gamma_0$ and $\zeta_0$, the non-equilibrium pressure can be found analytically
\begin{equation}
\Pi(r) = \Pi(r_*) \left(\frac{r_*}{r}\right)^{\frac23} + \frac{r}{15 \zeta_0 \gamma_0} \left[1- \left(\frac{r_*}{r}\right)^{\frac53}\right]\,.
\label{433}
\end{equation}
This function is non-monotonic, and for the appropriate values of the parameters $\Pi(r_*)$, $r_*$, $\zeta_0$ and $\gamma_0$ can change the sign and reach a minimum.

\section{The study of modified equation of hydrostatic equilibrium}

\subsection{The canonic equation of hydrostatic equilibrium}

When $q^i=0$, $\Pi^{ik}_{(0)}=0$, $\Theta =0$, and all the state functions depend on the radial variable only,  we obtain that the equation of the energy balance (\ref{05}) becomes trivial, and the equation (\ref{06}) takes the form
\begin{equation}
(W+{\cal P}) \ DU_k = \nablab_k {\cal P}  \,,
\label{35}
\end{equation}
where the total pressure  ${\cal P} = P-\frac13 \Pi$ contains both equilibrium and non-equilibrium parts.
Convolution of this equation with the director ${\cal R}^k$ yields
\begin{equation}
\frac{B^{\prime}}{B} = -  \frac{{2\cal P}^{\prime}}{(W+{\cal P})} \,.
\label{351}
\end{equation}
As usual, to find the functions $A(r)$ and $B(r)$ we address to the pair of Einstein's equations
\begin{equation}
\frac{1}{r^2} + \frac{A^{\prime}}{rA^2} - \frac{1}{r^2 A} = 8\pi G W \,,
\label{Ein1}
\end{equation}
\begin{equation}
-\frac{1}{r^2} + \frac{B^{\prime}}{rAB} + \frac{1}{r^2 A} = 8\pi G {\cal P} \,.
\label{Ein2}
\end{equation}
From the equation (\ref{Ein1}) one obtains immediately the standard solution
\begin{equation}
\frac{1}{A} = 1- \frac{2G M(r)}{r} \,,  \quad M(r) = 4\pi \int_0^r {r'}^2 dr' W(r') \,.
\label{Ein3}
\end{equation}
Using (\ref{351}) we obtain from (\ref{Ein2}) the equation of hydrostatic equilibrium
\begin{equation}
- r^2 {\cal P}^{\prime} =  \frac{G (W+{\cal P})(M + 4\pi r^3 {\cal P})}{\left(1-\frac{2GM}{r}\right)} \,.
\label{EHE}
\end{equation}
This integro-differential equation can be rewritten as the nonlinear differential equation of the second order
\begin{equation}
\left[\frac{r^2 {\cal P}^{\prime}+ 4\pi G r^3 {\cal P}(W+ {\cal P})}{2r {\cal P}^{\prime}-(W+{\cal P})} \right]^{\prime} =  4\pi G r^2 W \,.
\label{EHE2}
\end{equation}

\subsection{Equation of state}

We assume that the model of stellar configuration under study contains two constituents. The first one is the relativistic gas (fluid); the second element of the system is the radiation, which is in equilibrium with the radiative gas.
The total pressure of the system is assumed to be presented by the following function:
\begin{equation}
{\cal P} = n k_{\rm B}T + \frac13 \sigma_{(\rm SB)} T^4 - \frac13 \Pi \,.
\label{EoS3}
\end{equation}
Here $k_{\rm B}$ is the Boltzmann constant; $\sigma_{(\rm SB)}$ is the Stefan-Boltzmann constant equal to
\begin{equation}
\sigma_{(\rm SB)}={\frac {\pi ^{2}k_{\rm B}^{4}}{60c^{2}\hbar ^{3}}}\,.
\label{SB}
\end{equation}
For the total energy-density of the system we use the formula
\begin{equation}
W = n \left[m \frac{K_3(\lambda)}{K_2(\lambda)} - k_{\rm B} T \right] + \left(\sigma_{(\rm SB)} T^4 - \Pi \right)\,.
\label{EoS2}
\end{equation}
Here $n$ is the particle number density; $K_n(\lambda)$ are the McDonald functions given by
\begin{equation}
K_n(\lambda) = \frac{\lambda^n}{1 \cdot 3 \cdot \cdot \cdot (2n-1)} \int_0^{\infty} dt  e^{-\lambda \cosh{t}} \cdot \sinh^{2n}{t} \,,
\label{EHE33}
\end{equation}
with $\lambda= \frac{m}{k_{\rm B}T}$ (see the book \cite{deGroot} for details).

The pair of equations (\ref{EoS3}) and (\ref{EoS2}) gives us the specific version of the equation of state; one has to add to these equations the equation for the non-equilibrium pressure $\Pi$ (\ref{271}), and the equation for the temperature distribution (\ref{31}). Then, we put $W$ from (\ref{EoS2}) and ${\cal P}$ from (\ref{EoS3}) into the equation of hydrostatic equilibrium (\ref{EHE}) or (\ref{EHE2}) and obtain the key equation for the profile $n(r)$. Using the solution for $n(r)$ we can recover the state functions $W(r)$ and ${\cal P}(r)$ and then can reconstruct the metric functions $A(r)$ and $B(r)$. In general case this procedure can be realized only numerically, and we hope to fulfil such a detailed analysis in the next work. Below we consider, as an example, only  one specific solution for the toy model.

Mention should be made that such a representation of the equations of state for two-component relativistic system is disputable. The question is whether we can add or not the term $- \Pi$ into the expression for the total energy density (\ref{EoS2})? Our ansatz is that the term $\left(\frac13 \sigma_{(\rm SB)} T^4 {-} \frac13 \Pi \right)$ naturally appeared in the total pressure (\ref{EoS3}) relates to the radiation, and this term is equal to one third of the radiation energy density, appeared in (\ref{EoS2}). Similar problem appeared in the medium electrodynamics, when the terms containing both: the medium velocity and the terms connected with the electromagnetic field, should be classified and packed either to the stress-energy tensor of the electromagnetic field, or to the one of the matter. We think the formulas (\ref{EoS3}) and (\ref{EoS2}) form a special ansatz, and it has to be verified in the future.

\subsection{The example of exact solutions for a toy-model}

We assume now that the gas (fluid) is ultrarelativistic, i.e., $\lambda<<1$,  and its energy-density takes the form $W_{(\rm gas)} \approx 3k_{\rm B} n T$. Thus, the system as a whole is ultrarelativistic with $W=3{}\cal P$. For this model the equation of hydrostatic equilibrium is known to have a specific exact solution with
\begin{equation}
W(r) = \frac{W_0}{r^2} = 3 {\cal P} \,, \quad W_0 = \frac{3}{56\pi G} \,, \quad M(r)= \frac{3r}{14 G} \,, \quad  B(r) = B_0 \ r \,,  \quad A(r) = \frac74  \,.
\label{exact1}
\end{equation}
The spacetime with these metric coefficients is known to have conical singularity at the center, since $A(0)\neq 1$ and thus the Ricci scalar diverges $R(0)=\infty$ (see, e.g., the problem 16.13 in \cite{Problem}). Now we are ready to present analytically the profiles of the non-equilibrium pressure, temperature and particle number density.

\subsubsection{The model with $\gamma=0$, $\alpha_0 =0$ and $\tau_0<0$}

For this case we obtain that the non-equilibrium pressure and temperature have, respectively, the following form:
\begin{equation}
\Pi(r) = \Pi(r_*)   \left(\frac{r_*}{r}\right)^{\frac54}  \ \exp{\left(\frac{r_*-r}{9\zeta_0 |\tau_0|}\right)} \,, \quad T(r) = T(r_*)\left( \frac{r_*}{r}\right)^{\frac12} \,.
\label{2e}
\end{equation}
The particle number density can be found from the equation
\begin{equation}
W = 3 k_B n T + \sigma_{(\rm SB)} T^4 - \Pi \,.
\label{exact4}
\end{equation}
Now we obtain
\begin{equation}
n(r) = n(r_*)\left(\frac{r_*}{r} \right)^{\frac32} \left\{1+ \frac{\Pi(r_*)}{3k_{\rm B}T(r_*) n(r_*)}\left[\left(\frac{r}{r_*} \right)^{\frac14}\exp{\left(\frac{r_*-r}{9\zeta_0 |\tau_0|}\right)} -1 \right] \right\} \,.
\label{3e}
\end{equation}
The parameters $T(r_*)$, $\Pi(r_*)$ and $n(r_*)$ are linked by one relationship
\begin{equation}
\frac{3}{56\pi G r^2_*} = 3 k_B n(r_*) T(r_*) + \sigma_{(\rm SB)} T^4(r_*) - \Pi(r_*) \,,
\label{exact6}
\end{equation}
two of them should be chosen based on some physical assumptions.
Similarly to $W(r)$ and ${\cal P}(r)$, the solutions (\ref{2e}) and (\ref{3e}) are singular at the center and vanish at the infinity.

\subsubsection{The model with $\tau=0$}

Integration of the equation (\ref{33}) with $\gamma(r) = \gamma_0=const$ and $\zeta(r)T= \zeta_0 =const$ yields
\begin{equation}
\Pi(r) = \Pi(r_*)\left(\frac{r_*}{r} \right)^{\frac56} + \frac{\sqrt7 r}{33 \zeta_0 \gamma_0} \left[1-\left(\frac{r_*}{r} \right)^{\frac{11}{6}} \right]  \,.
\label{exact2}
\end{equation}
Integration of the equation (\ref{31}) with $\alpha_* = \frac{\alpha_0}{T}= const$ gives the following profile of the inverse temperature:
\begin{equation}
\frac{1}{T(r)} = \left(\frac{r}{r_*}\right)^{\frac12} \left\{\frac{1}{T(r_*)} + \frac{5\alpha_*}{8}\left[1-\left(\frac{r_*}{r}\right)^{\frac43} \right]
\left[\Pi(r_*) - \frac{\sqrt7 r_*}{33 \zeta_0 \gamma_0} \right] + \frac{2\sqrt7 \alpha_* r_*}{33 \zeta_0 \gamma_0}\left[1-\left(\frac{r}{r_*}\right)^{\frac12} \right] \right\}\,.
\label{exact3}
\end{equation}
Finally, we can formally write the particle number density in the form
\begin{equation}
n(r) = \frac{1}{3 k_B T}\left[\frac{3}{56\pi G r^2} - \sigma_{(\rm SB)} T^4 + \Pi \right] \,,
\label{exact5}
\end{equation}
where $T(r)$ and $\Pi(r)$ can be taken from (\ref{exact3}) and (\ref{exact2}), respectively.

\section{Conclusions}

1. We presented a new extended version of the relativistic non-equilibrium thermostatics. Terminologically, this extended theory can not be indicated as {\it causal thermostatics}, since standardly the causality of the thermodynamic processes is associated with the hyperbolic law of the heat propagation. However,  the extended formalism of the developed theory inherits the ideas,  on which the causal relativistic thermodynamics has been constructed in the works of Israel and Stewart, and in addition to the timelike unit medium velocity four-vector, the key element of the dynamic theory, the spacelike unit four-vector indicated as the {\it director}, is introduced into the static theory.

2. The established formalism is applied to the model of static spherically symmetric stellar structures, in the formation of which the radiation pressure plays the key role.  For this model the extended formalism gives a recipe how to calculate the non-equilibrium pressure and how to obtain the profile of the temperature with respect to the radial variable. Master equations of the model are based on the standard equation of hydrostatic equilibrium and are supplemented by the extended equations for the non-equilibrium pressure and temperature.

3. In order to illustrate the formalism, we analyzed in detail one toy-model corresponding to the ultrarelativistic state of matter interacting with radiation. We realize that this theory requires a multi-sectorial  numerical modeling, which we plan to consider in future investigations. The main expected result is the estimation of the star radius, which is defined as the first null of the generalized total pressure of the stellar configuration, and is the function of the set of the model guiding parameters.

\acknowledgments{The work was supported by Russian Science Foundation (Grant No 21-12-00130).}

\end{document}